\documentclass[11pt]{article}
\usepackage{fullpage}
\usepackage{appendix}
\usepackage[sort, authoryear, round]{natbib}
\usepackage{fancyhdr}
\usepackage[format=plain,
            labelfont={it,it},
            textfont=it]{caption}
\usepackage[stable, bottom, flushmargin]{footmisc}
\usepackage{graphicx}
\usepackage{xcolor}    
\usepackage{breakcites}
\usepackage{amsmath,amssymb,amsthm}
\usepackage{thmtools}
\usepackage{mathrsfs}
\usepackage{nicefrac}

\definecolor{lightblue}{RGB}{0,102,204}
\usepackage[pagebackref, colorlinks=true, linkcolor=black, citecolor=lightblue, urlcolor=lightblue]{hyperref}

\usepackage{pgfplots}
\pgfplotsset{compat = newest}
\usepackage{wrapfig}

\usepackage{enumitem}
\setlist[enumerate,1]{label=(\roman*)}
\usepackage{xparse, expl3}
\usepackage{indentfirst}
\usepackage{float}
\usepackage{tikz}
\usepackage{tikzscale}
\usetikzlibrary{arrows.meta,arrows}
\usetikzlibrary{decorations.pathreplacing}
\usetikzlibrary{positioning,chains,fit,shapes,calc,shapes.misc}

\usepackage[capitalize]{cleveref}

\newtheorem{theorem}{Theorem}[section]

\newtheorem{lemma}[theorem]{Lemma}
\newtheorem{corollary}[theorem]{Corollary}
\newtheorem{definition}[theorem]{Definition}

\newtheorem{conjecture}[theorem]{Conjecture}

\newtheorem{fact}[theorem]{Fact}

\crefname{theorem}{Theorem}{Theorems}
\crefname{proposition}{Proposition}{Propositions}
\crefname{lemma}{Lemma}{Lemmas}
\crefname{corollary}{Corollary}{Corollaries}
\crefname{definition}{Definition}{Definitions}
\crefname{example}{Example}{Examples}
\crefname{remark}{Remark}{Remarks}
\crefname{algorithm}{Algorithm}{Algorithms}
\crefname{equation}{Equation}{Equations}
\crefname{section}{Section}{Sections}
\crefname{subsection}{Section}{Sections}
\crefname{conjecture}{Conjecture}{Conjectures}
\crefname{fact}{Fact}{Facts}

\usepackage{microtype}
\usepackage{hyperref}
\usepackage{graphicx}

\usepackage{bbm,bm}
\usepackage{xfrac}
\usepackage{algorithm}
\usepackage{algpseudocode}
\usepackage{float}
\usepackage{mathtools}
\newcommand{\nc}{\newcommand}
\newcommand{\rnc}{\renewcommand}

\nc{\R}{\mathbb R}
\nc{\C}{\mathbb C}
\nc{\Q}{\mathbb Q}
\nc{\Z}{\mathbb Z}
\nc{\N}{\mathbb N}
\nc{\F}{\mathbb F}
\nc{\D}{\mathbb D}
\nc{\ind}[1]{\mathbb{I}[#1]}

\let\E\relax
\DeclareMathOperator*{\E}{\mathbb{E}}
\let\P\relax
\DeclareMathOperator*{\P}{\mathbb{P}}
\let\Pr\relax
\DeclareMathOperator*{\Pr}{\mathbb{P}}

\DeclareMathOperator*{\Var}{Var}

\nc{\CH}{\mathcal H}
\nc{\CD}{\mathcal D}
\nc{\CZ}{\mathcal Z}
\nc{\CP}{\mathcal P}
\nc{\CF}{\mathcal F}
\nc{\CG}{\mathcal G}
\nc{\CN}{\mathcal N}
\nc{\CX}{\mathcal{X}}
\nc{\CY}{\mathcal{Y}}
\nc{\CU}{\mathcal{U}}
\nc{\CA}{\mathcal{A}}
\nc{\CE}{\mathcal{E}}
\nc{\CI}{\mathcal{I}}
\nc{\CC}{\mathcal{C}}
\nc{\CR}{\mathcal{R}}
\nc{\CS}{\mathcal{S}}
\nc{\CB}{\mathcal{B}}

\nc{\VC}{\mathrm{VC}}
\nc{\DS}{\mathrm{DS}}
\nc{\Nat}{\mathrm{Nat}}
\nc{\Graph}{\mathrm{Graph}}

\nc{\defn}[1]{\textit{#1}}
\nc{\eg}{\emph{e.g.,~}}
\nc{\ie}{\emph{i.e.}}
\nc{\ER}{Erd\H{o}s-R\'enyi}

\rnc{\t}[1]{\text{#1}}

\nc{\SG}{\mathscr{G}}
\nc{\eps}{\varepsilon}
\nc{\on}[1]{\operatorname{#1}}
\nc{\Aut}{\on{Aut}}
\newcommand\restr[2]{{
  \left.\kern-\nulldelimiterspace
  #1 
  \vphantom{\big|}
  \right|_{#2}
  }}

\DeclarePairedDelimiter{\floor}{\lfloor}{\rfloor}
\nc{\ol}[1]{\overline{#1}}
\nc{\wt}[1]{\widetilde{#1}}
\nc{\vp}{\varphi}
\renewcommand{\ln}{\log}
\nc{\ev}{E}
\nc{\p}{\textbf{\textit{p}}}

\newtheorem{observation}[theorem]{\textbf{Observation}}

\title{Semi-Random Graphs, Robust Asymmetry, and Reconstruction}

\author{
Julian Asilis\footnote{University of Southern California. Email: \href{mailto:asilis@usc.edu}{asilis@usc.edu}. Supported by the National Science Foundation Graduate Research Fellowship
Program under Grant No.\ DGE-1842487.} \and 
Xi Chen\footnote{Columbia University. Email: \href{mailto:xc2198@columbia.edu}{xc2198@columbia.edu}.
Supported by NSF awards CCF-2106429 and CCF-2107187.} \and 
Dutch Hansen\footnote{University of Southern California. Email: \href{mailto:jmhansen@usc.edu}{jmhansen@usc.edu}.}  \and 
Shang-Hua Teng\footnote{University of Southern California. Email: \href{mailto:shanghua@usc.edu}{shanghua@usc.edu}. Supported by Simons Foundation Investigator Award.}
}

\date{}

\begin{document}

\maketitle

\begin{abstract}
The Graph Reconstruction Conjecture famously posits that any undirected graph on at least three vertices is determined up to isomorphism by its family of (unlabeled) induced subgraphs. At present, the conjecture admits partial resolutions of two types: 1) casework-based demonstrations of reconstructibility for families of graphs satisfying certain structural properties, and 2) probabilistic arguments establishing reconstructibility of random graphs by leveraging average-case phenomena. While results in the first category capture the worst-case nature of the conjecture, they play a limited role in understanding the general case. Results in the second category address much larger graph families, but it remains unclear how heavily the necessary arguments rely on optimistic distributional properties. Drawing on the algorithmic notions of smoothed and semi-random analysis, we study the robustness of what are arguably the two most fundamental properties in this latter line of work: asymmetry and uniqueness of subgraphs. Notably, we find that various natural \emph{semi-random} graph distributions exhibit these properties asymptotically, much like their Erd\H{o}s-R\'enyi counterparts.

In particular, \citet{bollobas1990almost} demonstrated that almost all Erd\H{o}s-R\'enyi random graphs $G = (V, E) \sim \mathscr{G}(n, p)$ enjoy the property that their induced subgraphs on $n - \Theta(1)$ vertices are asymmetric and mutually non-isomorphic, for $1 - p, p = \Omega(\log(n) / n)$. As our primary result, we demonstrate that this property is \emph{robust} against perturbation---even when an adversary is permitted to add/remove each vertex pair in $V^{(2)}$ with (independent) arbitrarily large constant probability. Exploiting this result, we derive asymptotic characterizations of asymmetry in random graphs with large planted structure and bounded adversarial corruptions, along with improved bounds on the probability mass of nonreconstructible graphs in $\mathscr{G}(n, p)$.
\end{abstract}

\clearpage

\section{Introduction}\label{Section:Introduction}

The Graph Reconstruction Conjecture, or simply the Reconstruction Conjecture, dates to the work of \citet{ulam1960collection} and \citet{kelly1942isometric}, and remains one of the foremost open problems in graph theory. Informally, the conjecture states that each undirected graph on at least three vertices is \emph{reconstructible}, meaning it is determined up to isomorphism by its multiset of unlabeled induced subgraphs. 

Since its proposal eight decades ago, numerous partial results have demonstrated the reconstructibility of graphs under various restricted conditions. Broadly speaking, these results --- and the techniques used to establish them --- typically fall into one of two categories: 1) proofs of reconstructibility for families of graphs obeying certain structural properties, via highly bespoke casework, and 2) proofs of ``almost everywhere'' reconstructibility for certain standard graph distributions (\eg \ER), which exploit the asymptotic behavior of such distributions. 
While results in the first category capture the worst-case spirit of the Reconstruction Conjecture, guaranteeing reconstructibility for all graphs within a certain family, such families can be restrictive, and the associated techniques rarely extend to new settings. Results in the second category address much larger graph families, but typically rely upon brittle properties of random graphs (\eg asymmetry), which are probabilistically common but often not enjoyed by graphs encountered in practice. 

In light of these limitations, we study the Graph Reconstruction Conjecture from a flexible probabilistic perspective which smoothly interpolates between the analysis of specialized graph distributions, which experienced rapid success in the establishment of strong theorems, and the classic form of the conjecture, which has remained notoriously resistant to proof. In particular, we direct our attention to the following two questions: 

\begin{enumerate}
    \item[(\emph{i.})] \emph{Given a fixed underlying graph $G$, how much random noise must be applied to $G$'s edges in order to ensure that the resulting graph is reconstructible with high probability?}
    \item[(\emph{ii.})] \emph{How robust is the reconstructibility of random graphs? What if an adversary is permitted to modify the random graph, under certain restrictions, before its reconstructibility is tested?}
\end{enumerate}

Note that Question ($i.$) commences with the worst-case perspective of the Reconstruction Conjecture by fixing an underlying graph $G$ --- thought of as adversarially selected --- yet mollifies some of its structure by way of random edge noise. Patently, this perspective is motivated by the smoothed analysis of \cite{spielman2004smoothed}, which measures the complexity of algorithms on worst-case inputs that experience slight random perturbations before  their runtimes are measured. In the context of graph reconstruction, this outlook can be thought of as measuring the brittleness or sparsity of nonreconstructible graphs; in order for an adversary to succeed, they must select a graph whose nonreconstructibility is unlikely to be ameliorated by edge noise. 

By a similar token, Question ($ii.$) modifies the standard probabilistic treatment of graph reconstruction by endowing it with a component of worst-case analysis. Namely, an adversary is granted the opportunity to adjust each random graph before examining its reconstructibility, potentially allowing them to destroy brittle graph properties previously used to demonstrate reconstrucibility (\eg subgraph asymmetry). This perspective is inspired by the analysis of algorithms in the semi-random setting \citep{blum2995coloring,feige2001heuristics}, in which inputs are drawn according to a natural distribution and subsequently modified by an adversary with constrained editing power. 

Drawing on both of these techniques, we introduce a semi-random perspective on the reconstruction problem, living between deterministic and probabilistic settings. From this vantage point, we seek to characterize the reconstructibility of large graph families in a manner that admits structural irregularity. While we do not introduce novel techniques for the deterministic conjecture, we find that natural semi-random graph distributions exhibit surprising asymmetry properties, much like their \ER{} counterparts, which may be of independent interest. As a consequence in our setting, we observe that existing deterministic reconstruction arguments are much stronger than previously recognized.

\subsection{Primary Results and Technical Overview}\label{sec:primary-results}

For a fixed graph $G_0$ and $p \in (0, 1)$, consider the random graph $G$ given by independently flipping each edge/nonedge in $G_0$ with independent probability $p$. Note that this is equivalent to independently including edges that do not appear in $G_0$ with probability $p$ and edges that do appear in $G_0$ with probability $1-p$. Moreover, if $p \in O(1/n^2)$, $n$ can be taken sufficiently large so that $G = G_0$ with probability $\Omega(1)$. This observation suggests one possible measure of progress towards the full conjecture: the smallest $p(n)$ such that graphs perturbed in the manner remain reconstructible asymptotically.

We encapsulate this smoothed perspective by considering a generalization of $\SG(n, p)$ which allows for non-uniform edges probabilities. Let $n \in \N$ and $N = \binom{n}{2}$. For a vertex set $V$, $|V| = n$, we fix an arbitrary indexing $\pi \colon V^{(2)} \leftrightarrow [N]$. For $\p = (p_i)_i \in [0,1]^N$, we denote by $\SG(n, \p)$ the distribution over graphs on $V$ such that each vertex pair $e \in V^{(2)}$ is included as an edge with independent probability $p_{\pi(e)}$. Equivalently, $\SG(n, \p) = \bigotimes_{1}^{N} \on{Ber}(p_i)$. Of course, we will consider probabilities that vary with $n$; in full generality, we let $\p : \N \to [0,1]^*$ such that $\p(n) \in [0,1]^{N}$. In this case, we denote by $\SG(n, \p)$ the distribution given by including an edge $e$ with probability equal to the $\pi(e)$th coordinate of $\p(n)$.

Employing the semi-random perspective, we have some flexibility. For a graph $G = (V, E)$ and subset $S \subset V^{(2)}$, we denote by $B_S(G)$ the family of graphs generated by adding/removing any number of vertex pairs from $S$ in $G$. That is, 
\[B_S(G) = \left\{G + \sum_{e \in S} c_e e : c_e \in \{\pm 1\} \right\}.\] 
For $r \in \N$, we denote by $B_r(G)$ the family of graphs generated by adding/removing at most $r$ vertex pairs anywhere in $G$,
\[B_r(G) = \bigcup_{|S| \leq r} B_S(G).\] 
Intuitively, should we allow an adversary to edit a limited number of edges anywhere in $G$, we hope that all graphs in $B_\eps(G)$ be reconstructible. Should we constrain an adversary's changes to a (possibly random) set of pairs $S$, we hope that all graphs in $B_S(G)$ be reconstructible. To start, we find it most fruitful to uniformly draw a subset $S \subset V^{(2)}$ of order $\eps = \eps(n)$ and examine the behavior of $B_S(G)$. Note that for appropriate $\eps \in \Omega(n^2 \log(n))$, an asymptotic reconstructibility result for $B_S(G)$ would suffice to prove the Reconstruction Conjecture for sufficiently large $n$ (see \cref{sec:deterministic-reconstruction}). This observation furnishes another measure of progress towards the full conjecture: the largest $\eps(n)$ such that all graphs in $B_S(G)$ remain reconstructible asymptotically. Loosely, this formalization of the semi-random model can be seen as, instead of uniformly widening the hypercube bound on $\p \in [0,1]^N$ (as is our objective under the smoothed interpretation), allowing a random (or adversarially chosen) subset of the entries in this tuple to be arbitrary. Studying these settings, we give strengthened and refined forms of existing probabilistic graph reconstruction results. We summarize our contributions as follows.

\paragraph{Robust graph asymmetry.} Our primary technical contribution characterizes the asymmetry and uniqueness of large subgraphs in $\SG(n, p)$ under randomly constrained edge perturbations.
\begin{theorem}[Informal \cref{thm:bollobas-generalization-wo-replacement,thm:bollobas-generalization-wo-replacement-constant-prob}]\label{thm:bollobas-generalization-wo-replacement-informal}
    Let $\delta \in \N$, $c \in (0,1/2)$, $\beta : \N \to (0,1)$, and $\eps : \N \to \N$ such that $1-\beta, \beta \in \Omega(\log(n)/n)$ and $\eps = \eps(n) \leq cn^2$. There exists $\alpha > 0$ such that if $\p(n) \in [\alpha \beta(n), 1-\alpha \beta(n)]^{\binom{n}{2}}$, then for $G \sim \mathscr{G}(n, \p)$ and independent uniformly drawn subset $S \subset V^{(2)}$ of order $\eps$, it holds with failure probability at most $\exp(-\Omega(\beta(n)n))$ that for all $H \in B_S(G)$, all induced subgraphs of $H$ on $n-\delta$ vertices are asymmetric and mutually non-isomorphic. 
\end{theorem}

\noindent We remark that for $\beta(n) \in o(1)$ and fixed $\rho > 0$, we can take $\alpha > 0$ so that the failure probability is bounded by $\exp(-\rho\beta(n)n)$. Up to constants in the bound on $p(n)$, the asymmetry result of \citet{bollobas1990almost} can be seen as the special case where $\eps = 0$.

\paragraph{Semi-random graph asymmetry.} When $G \sim \SG(n, 1/2)$ and no edits are allowed, enumerative techniques similar to those of \citet{muller1977edge} give a failure probability in \cref{thm:bollobas-generalization-wo-replacement-informal} of $\exp(-\Omega(n))$. If we want to ensure asymmetry on all graphs $G' \in B_\eps(G)$, we can simply observe an $O(n^{2\eps})$ cardinality blowup, so it is enough to take $\eps \leq Bn/\log(n)$ for carefully chosen $B>0$. This does not translate nicely to the non-uniform setting, however (\textit{e.g.} $p \neq 1/2$). Using \cref{thm:bollobas-generalization-wo-replacement-informal}, we obtain a strict generalization of this observation via conditioning. We obtain even stronger results when adversarial perturbations are restricted to (or planted on) fixed subsets of $V^{(2)}$.

\begin{corollary}[Informal \cref{cor:wo-replacement-small-p-planted,cor:wo-replacement-constant-p-planted}]\label{cor:wo-replacement-planted-informal}
    Let $\delta \in \N$ and $\beta : \N \to (0,1)$ such that $1-\beta, \beta \in \Omega(\log(n)/n)$. There exist $\alpha, B > 0$ such that if $\p(n) \in [\alpha \beta(n), 1-\alpha \beta(n)]^{\binom{n}{2}}$, then for $G \sim \mathscr{G}(n, \p)$ and any fixed $P \subset V^{(2)}$ such that $|P| \leq B\beta(n)n$, it holds with failure probability at most $\exp(-\Omega(\beta(n)n))$ that for all $H \in B_P(G)$, all induced subgraphs of $H$ on $n-\delta$ vertices are asymmetric and mutually non-isomorphic.
\end{corollary}

\noindent Here and in the remainder of the overview, if $\beta \in o(1)$ then $\rho, B > 0$ are free parameters; we may choose $\alpha > 0$ so that our results holds for a chosen $B > 0$ with failure probability at most $\exp(-\rho\beta(n)n)$. 

For $|\p|_\infty \in \Theta(1)$, our result is tight up to constants; trivially, one can plant $2n -3$ nonedges to guarantee the existence of a chosen transposition on any sampled graph. As a concrete illustration, when $p \in \Theta(1)$, one preserves asymmetry even when a clique is planted on $\Theta(\sqrt{n})$ vertices. In contrast, the clique number of $\mathscr{G}(n, p)$ is known to be $\Theta(\log n)$ asymptotically almost surely \citep{grimmett1975colouring}. 

\begin{corollary}[Informal \cref{cor:wo-replacement-small-p-ball,cor:wo-replacement-constant-p-ball}]\label{cor:wo-replacement-p-ball-informal}
    Let $\delta \in \N$ and $\beta : \N \to (0,1)$ such that $1-\beta, \beta \in \Omega(\log(n)/n)$. There exist $\alpha, B > 0$ such that if $\p(n) \in [\alpha \beta(n), 1-\alpha \beta(n)]^{\binom{n}{2}}$ and $\eps : \N \to \N$ satisfies $\eps(n) \leq B\beta(n)n/\log(n)$, then for $G \sim \mathscr{G}(n, \p)$, it holds with failure probability at most $\exp(-\Omega(\beta(n)n))$ that for all $H \in B_\eps(G)$, all induced subgraphs of $H$ on $n-\delta$ vertices are asymmetric and mutually non-isomorphic. 
\end{corollary}

\noindent One can view \cref{cor:wo-replacement-p-ball-informal} as establishing robustness against perturbations made \textit{after} the graph has been sampled. Though \cref{cor:wo-replacement-p-ball-informal} admits logarithmically smaller $\eps$ than the results of \cite{erdos1963asymmetric,kim2002asymmetry}, and does not provide tight characterizations of resilience, we address the nonexistence of isomorphisms between \textit{proper} subgraphs and allow for non-uniformity in edge probabilities, which rules out exact computation.\footnote{Note that \cref{cor:wo-replacement-p-ball-informal} devolves into a global asymmetry result when $\delta = 1$; note that all subgraphs on $n-1$ vertices being mutually non-isomorphic is sufficient for global asymmetry.} This latter point enables the smoothed setting, in which edge probabilities may differ by $1-o(1)$. 

\paragraph{Probabilistic graph reconstruction.} While our asymmetry results may be of independent interest, the consequences in our setting are immediate. Taking $\delta = 2$, the conclusions of \cref{thm:bollobas-generalization-wo-replacement-informal,cor:wo-replacement-planted-informal,cor:wo-replacement-p-ball-informal} are sufficient for reconstructibility via \cref{cor:farhadian}. This yields reconstruction results for random graphs with: 1) adversarial edge edits constrained to a random subset of $O(n^2)$ vertex pairs, 2) planted structure on $O(\beta(n)n)$ fixed vertex pairs, or 3) at most $O(\beta(n)n/\log(n))$ adversarial edge edits anywhere in the graph. We state these results in full generality in \cref{sec:planted-structure}. After providing some notation, we state special cases here. 

For a graph $G = (V, E)$ and $F \subset V^{(2)}$, define $G + F = (V, E \cup F)$ and $G - F = (V, E\setminus F)$. Also let $\mathfrak{R}_n$ denote the family of reconstructible graphs on $n$ vertices, and $\ol{\mathfrak{R}}_n$ the family of nonreconstructible graphs on $n$ vertices. 

\begin{corollary}
    Let $\beta : \N \to (0,1)$ such that $1-\beta, \beta \in \Omega(\log(n)/n)$. There exist $\alpha, B > 0$ such that if $\p(n) \in [\alpha \beta(n), 1-\alpha \beta(n)]^{\binom{n}{2}}$, then for sufficiently large $n$, $G \sim \mathscr{G}(n, \p)$, and any fixed $P_1, P_2 \subset V^{(2)}$ such that $|P_1 \cup P_2| \leq B\beta(n)n$, it holds with failure probability at most $\exp(-\Omega(\beta(n)n))$ that $G + P_1 - P_2$ is reconstructible.
\end{corollary}

\begin{corollary}[Informal \cref{cor:nonrecon-balls-small-p,cor:nonrecon-balls-constant-p}]\label{cor:nonreconstructible-informal}
    Let $\beta : \N \to (0,1)$ such that $1-\beta,\beta \in \Omega(\log(n)/n)$. There exist $\alpha, B > 0$ such that if $\p(n) \in [\alpha\beta(n),1-\alpha\beta(n)]^{\binom{n}{2}}$ and $\eps \leq B\beta(n)n/\log(n)$, then \[\Pr_{\SG(n, \p)}[B_\eps(\ol{\mathfrak{R}}_n)] \xrightarrow{n\to\infty} 0.\]
\end{corollary}

\noindent When each edge is drawn with uniform probability $p \in \Theta(\log(n)/n)$, the result of \citet{bollobas1990almost} guarantees convergence to zero only for $\eps = 0$. \cref{cor:nonreconstructible-informal} demonstrates that in this same regime, we can in fact take any constant $\eps$. Using techniques from \citet{bollobas1990almost}, one can show that the same results hold if we instead let $\mathfrak{R}_n$ denote the family of graphs on $n$ vertices with reconstruction number three.\footnote{A graph is said to have reconstruction number $k \leq n$ if it determined by some order-$k$ collection of its induced subgraphs \citep{harary1985reconstruction}.}

\subsection{Related Work}
\label{subsec:related-work}

In the present work, we are concerned with the Vertex Reconstruction Conjecture. Graph families proven to be reconstructible in this setting include regular graphs, disconnected graphs, and trees \citep{kelly1957congruence}. More recently, \cite{farhadian2017simple} observed connections between the reconstructibility of a graph and the uniqueness and asymmetry of its large subgraphs. On the empirical front, \citet{mckay1997small,mckay2022reconstruction} verified that all graphs up to 13 vertices are reconstructible. For an overview of standard results and alternative variants of the conjecture, see \cite{bondy1991manual}. 

The probabilistic examination of graph reconstruction dates to \cite{muller1976probabilistic}, who proved that for any $\alpha \in (0, 1)$, almost all graphs are reconstructible from their multiset of subgraphs on $\frac{n}{2}(1+\alpha)$ vertices. Using an enumerative argument, M\"uller proved that for $G \sim \SG(n, 1/2)$, it holds with probability $1-o(1)$ that all sufficiently large subgraphs of $G$ must be asymmetric and mutually non-isomorphic, a sufficient condition for reconstructibility. The seminal work of \cite{bollobas1990almost} refined the analysis for $\SG(n, p)$ with $p \in \Omega(\log(n)/n)$, proving that asymptotically almost surely $G \sim \SG(n, p)$ is reconstructible from any three subgraphs on $n-1$ vertices. More recently, \cite{spinoza2019reconstruction} used the asymmetry result of \cite{muller1976probabilistic} to prove that for $\ell \leq \frac{n}{2}(1-\alpha)$, almost all graphs are reconstructible from some multiset of $\binom{\ell+2}{2}$ subgraphs on $n-\ell$ vertices.

The literature on graph asymmetry is even more extensive. \cite{erdos1963asymmetric} initiated the study of this notion by proving that almost all graphs are asymmetric. In particular, they proved that asymptotically almost surely, one must remove and add a total of at least $\frac{n}{2}(1-\alpha)$ edges to make $G \sim \SG(n, 1/2)$ symmetric. In recent years, this result has joined the literature on property resilience \citep{sudakov2008local,frieze2015introduction,bollobas2011random}. More generally, it is folklore that for $1-p, p \geq \log(n)/n$, one must asymptotically almost surely add and remove at least $(2 + o(1))np(1-p)$ edges somewhere in $G \sim \SG(n, p)$ to achieve symmetry. 
\cite{kim2002asymmetry} refined this result by proving that for the same range of $p$, one must asymptotically almost surely add and remove at least $(1 + o(1))np(1-p)$ edges adjacent to a single vertex. A number of works have also studied asymmetry in different graph distributions \citep{kim2002asymmetry, bollobas1982asymptotic, mckay1984automorphisms, odonnell2014hardness}. The work of \cite{odonnell2014hardness} comes closest to ours in terminology; examining hardness in a robust variant of the Graph Isomorphism problem, they prove that the uniform distribution over graphs on $n$ vertices and $m \in \Omega(n)$ edges satisfies a strengthened notion of global asymmetry that decays with $n$ if $m$ is super-linear. 

Also related to the sensitivity of graph asymmetry is the work of \cite{brewer2018asymmetric} and \cite{hartke2009automorphism}. These works examine changes in the graph automorphism group under addition and deletion of carefully chosen edges. \cite{brewer2018asymmetric} take on a dual perspective to that of \cite{erdos1963asymmetric}, studying the minimum number of edges one must add and delete in order to make a graph asymmetric. \cite{hartke2009automorphism} prove that for any two finite groups, there exists a graph from which one can remove a vertex so that the automorphism group passes between these groups. This result can be seen as a barrier to proving the Reconstruction Conjecture via characterization of subgraph symmetries.

\section{Preliminaries}\label{Section:Preliminaries}

\subsection{Notation}

For $r \in \R$, we let $\floor{r}$ denote the greatest integer $m \leq r$. For $m \in \N$, we let $[m] = \{1, 2,\ldots,m\}$. For a finite set $U$ and natural numbers $k \leq |U|$, $\ell \leq \binom{|U|}{k}$, we let $U^{(k)}$ denote the set of all $k$-element subsets of $U$, and $U^{(k, \ell)} = U^{(k)(\ell)}$ denote the set of all $\ell$-element subsets of $U^{(k)}$. For a random variable $X$ and event $E$, we let $\Pr_X$ denote the distribution of $X$ and $\ind{E}$ denote the indicator of $E$. For a finite set $U$, we let $\CU(U)$ denote the uniform distribution over $U$. For any distribution $\CD$ supported on $U$ and $k \in \N$, we let $\CD^{\otimes k}$ denote the $k$-fold product of $\CD$, supported on $U^k$.

\subsection{Graph Reconstruction}

We direct our attention exclusively to finite undirected graphs; for an introduction to standard notation, see \cite{bollobas2002modern, bollobas2011random}.
For a graph $G = (V, E)$, we denote $v(G) = |V|$ and $e(G) = |E|$. For a set of vertices $U \subseteq V$, $G[U]$ denotes the \defn{induced subgraph} of $G$ on $U$, \ie, $G[U] = (U, \{e \in E : e \subset U\})$.
For a vertex $v \in V$, we let $G - v = G[V\setminus\{v\}]$. More generally, for any subset $U \subset V$, we let $G - U = G[V\setminus U]$. Analogously, for any subset $F \subset V^{(2)}$, we let $G + F = (V, E \cup F)$ and $G - F = (V, E\setminus F)$. Henceforth, for concision, we refer to induced subgraphs simply as subgraphs.

A \defn{graph isomorphism} $\varphi \colon G \xrightarrow{\cong} H$ is a bijection $\varphi \colon V(G) \to V(H)$ such that $xy \in E(G)$ if and only if $\vp(x)\vp(y) \in E(H)$. If there exists an isomorphism $\varphi \colon G \xrightarrow{\cong} H$, we say that $G$ and $H$ are \defn{isomorphic} and write $G \cong H$. A \defn{graph automorphism} on $G$ is an isomorphism $\varphi \colon G \xrightarrow{\cong} G$. We let $\Aut G$ denote the set of automorphisms on $G$. The identity is always an automorphism; if no other automorphism exists, we call $G$ \defn{asymmetric}. If $G$ is not asymmetric, we call it \defn{symmetric}. We say that two vertices $x, y \in V(G)$ are \defn{similar}, and write $x \sim y$, if there exists $\vp \in \Aut G$ such that $\vp(x) = y$. We say that a set of vertices $U \subset V(G)$ is \defn{fixed} in $G$ if $\vp(U) = U$ for all $\varphi \in \Aut G$. We say a subgraph of $G$ is \defn{unique} if it is isomorphic to no other subgraph of $G$. 

A \defn{hypomorphism} $\sigma \colon G \xrightarrow{\sim} H$ is a bijection $\sigma \colon V(G) \to V(H)$ such that $G - v \cong H - \sigma(v)$ for all $v \in V(G)$. If there exists a hypomorphism $\sigma \colon G \xrightarrow{\sim} H$, we say that $G$ and $H$ are \defn{hypomorphic} and write $G \sim H$. Naturally, $G \cong H$ implies $G \sim H$. The Reconstruction Conjecture asserts the converse, for graphs on at least three vertices.

\begin{conjecture}[\cite{ulam1960collection,kelly1942isometric}]
    Let $G, H$ be undirected graphs on at least $3$ vertices. If $G \sim H$, then $G \cong H$.
\end{conjecture}

\noindent In general, we say that a graph $G$ is \defn{reconstructible} if whenever $G \sim H$, it holds that $G \cong H$. A family of graphs $\CC$ is reconstructible if all $G \in \CC$ are reconstructible. We let $\mathfrak{R}_n$ denote the family of reconstructible graphs on $n$ vertices, and $\ol{\mathfrak{R}}_n$ the family of nonreconstructible graphs on $n$ vertices.
A classical tool in graph reconstruction is Kelly's lemma, which we state below alongside two relevant facts.
\begin{lemma}[\cite{kelly1957congruence}]
    Suppose $G \sim H$. Then for any graph $F$ on at most $n-1$ vertices, $G$ and $H$ have the same number of subgraphs isomorphic to $F$.
\end{lemma}

\begin{fact}\label{fact:hypomorphism-preserves-anchor}
    Let $\sigma \colon G \xrightarrow{\sim} H$, and $U \subset V(G)$ such that $G[U]$ is a unique subgraph of $G$. Then $G[U] \cong H[\sigma(U)]$, and $H[\sigma(U)]$ is unique in $H$.
\end{fact}

\begin{proof}
    Let $\vp_x : G - x \xrightarrow{\cong} H -\sigma(x)$ denote isomorphisms given by the existence of $\sigma$.
    By Kelly's lemma, there is a subset $U' \subset V(H)$ such that $H[U'] \cong G[U]$ and $H[U']$ is unique in $H$. Also note that for all $x \in V \setminus U$, $\vp_x$ restricts to an isomorphism $G[U] \xrightarrow{\cong} H[\vp_v(U)]$, so $H[\vp_v(U)]$ is a subgraph of $H$ isomorphic to $G[U]$, and $\vp_v(U) = U'$. In particular, $\sigma(x) \notin U'$ for all $x \in V \setminus U$, and $\sigma(U) = U'$.
\end{proof}

\section{Results}

\subsection{Asymmetry}

Asymmetry properties are commonly leveraged in probabilistic reconstruction arguments. Namely, such arguments often invoke the asymmetry of all sufficiently large (induced) subgraphs of the underlying random graph. 
The dependence of this technique upon all subgraphs enjoying this property can occasionally be avoided by instead assuming the \emph{uniqueness} of certain subgraphs.
\citet{farhadian2017simple} exploited this approach, demonstrating that all graphs containing a unique and asymmetric subgraph on $n - 2$ vertices are reconstructible.  
As our first lemma, we present a mild generalization of this result, which admits a constructive proof.

\begin{lemma}\label{lem:n-2-anch-generalization}
    Suppose there exists a hypomorphism $\sigma \colon G \xrightarrow{\sim} H$, vertices $x, y \in V(G)$, and isomorphisms 
    $\vp_x \colon G - x \xrightarrow{\cong} H - \sigma(x)$ and $\vp_y \colon G - y \xrightarrow{\cong} H - \sigma(y)$ such that \begin{enumerate}
        \item $\vp_x(y) = \sigma(y)$,
        \item $\vp_y(x) = \sigma(x)$,
        \item $S_x = \big\{v \in V(G)\setminus\{x, y\} : xv \in E(G) \big \}$ is fixed in $G - \{x, y\}$.
    \end{enumerate}
    Then $G \cong H$.
\end{lemma}

\begin{proof}
    For brevity, let $A = G - \{x, y\}$. Note that by conditions ($i$) and ($ii$), the restrictions $\restr{\vp_x}{A}, \restr{\vp_y}{A}$ are both isomorphisms $G - \{x, y\} \xrightarrow{\cong} H - \{\sigma(x), \sigma(y)\}$. Hence, we have \[\restr{\vp_x}{A}^{-1} \circ \restr{\vp_y}{A} \in \Aut(A).\] Condition ($iii$) then implies $\vp_x(S_x) = \vp_y(S_x)$. Now, because $\vp_y(x) = \sigma(x)$, 
    \[\restr{\vp_x}{A}(S_x) = \restr{\vp_y}{A}(S_x) = \bigg\{w \in V(H) \setminus \big\{\sigma(x),\sigma(y) \big \} : \sigma(x)w \in E(H)\bigg\}.\] 
    We can therefore extend $\vp_x$ to an isomorphism $\ol{\vp}_x : G \xrightarrow{\cong} H$ by defining \[\ol{\vp}_x(v) = \begin{cases}
        \sigma(x) &v = x\\
        \vp_x(v) &\text{else}.
    \end{cases}\]
    This mapping preserves edges and nonedges except for possibly $xy$. However, as $G \sim H$, then  $e(G) = e(H)$ and $xy \in E(G)$ if and only if $\ol{\vp}_x(x)\ol{\vp}_x(y) \in E(H)$.
\end{proof}

\begin{corollary}[{\cite{farhadian2017simple}}]\label{cor:farhadian}
    If $G$ has a unique, asymmetric subgraph on $n-2$ vertices, then $G$ is reconstructible.
\end{corollary}

\begin{proof}
    Let $G - \{x, y\}$ denote a subgraph of $G$ that is both unique and asymmetric. 
    Let $\sigma : G \xrightarrow{\sim} H$. By \cref{fact:hypomorphism-preserves-anchor}, there is a unique and asymmetric subgraph $H - \{x', y'\}$ of $H$ such that $H - \{x', y'\} = H[\sigma(V(G) \setminus \{x, y\})]$. Hence, $\{\sigma(x), \sigma(y)\} = \{x', y'\}$. 

    Let $\vp_x : G - x \xrightarrow{\cong} H - \sigma(x)$ and $\vp_y : G = y \xrightarrow{\cong} H - \sigma(y)$. By uniqueness of $H - \{x', y'\}$ in $H$, $\vp_x(V(G) \setminus \{x, y\}) = \vp_y(V(G) \setminus \{x, y\}) = V(H) \setminus \{x', y'\}$. So $\vp_x(y) = \sigma(y)$ and $\vp_y(x) = \sigma(x)$. Finally, note that condition \emph{(iii)} of \cref{lem:n-2-anch-generalization} holds by the assumption that $G - \{x, y\}$ is asymmetric.
\end{proof}

\subsection{Robust Asymmetry}\label{sec:robust-asymmetry}

Let us begin by articulating subgraph uniqueness properties that will play a central role in our forthcoming results. 
Recall that for a graph $G = (V, E)$ and subset $S \subset V^{(2)}$, $B_S(G)$ denotes the set of graphs generated by adding/removing any number of vertex pairs from $S$ in $G$,
\[ B_S(G) = \left\{G + \sum_{e \in S} c_e e : c_e \in \{\pm 1\} \right\}. \] 

\begin{definition}
Let $n, \delta \in \N$. Let $G = (V, E)$ be a random graph on $n + \delta$ vertices, $\CC$ be a random set of graphs on $V$, and $S$ be a random subset (or tuple)
of pairs in $V^{(2)}$. We let $\ev_\delta(G)$ denote the event that: if $W \subset V$, $|W| = n$ and $\varphi : W \to V$ is an injection inducing a graph isomorphism $G[W] \xrightarrow{\cong} G[\varphi(W)]$, then $\varphi(w) = w$ for all $w \in W$. Overloading notation, we also let \[\ev_\delta(\CC) = \bigwedge_{G \in \CC} \ev_\delta(G), \ \ \ \ \ev_\delta(G, S) = \ev_\delta(B_S(G)).\]
\end{definition}

Informally, $E_{\delta}(G)$ records the condition that all subgraphs of $G$ on $n$ vertices are asymmetric and mutually non-isomorphic, while $E_{\delta}(G, S)$ records whether this is true even when one is permitted to add/remove any vertex pairs in $S$ from $G$. Note that the negation of $E_\delta(G, S)$ is monotonic in $S$.
We are now equipped to state our primary lemma.

\begin{lemma}\label{lem:bollobas-generalization}
Let $\eps = O(n^2)$, $\delta \in \N$, $\rho > 0$. There exists $\alpha = \alpha_{\eps, \delta, \rho} > 0$ such that if 
\[\frac{\ln(n)}{n} \leq \beta(n) = o(1), \qquad \quad \p(n) \in [\alpha\beta(n), 1-\alpha\beta(n)]^{\binom{n}{2}} \]
then for sufficiently large $n$, $G \sim \mathscr{G}(n + \delta, \p)$, and independent $S \sim \CU(V^{(2)})^{\otimes \eps}$, $\ev_\delta(G, S)$ holds with failure probability at most $\exp(-\rho \beta(n)n)$. 
\end{lemma} 

\begin{lemma}\label{lem:bollobas-generalization-constant-prob}
    Let $\eps = O(n^2)$, $\delta \in \N$, $\beta \in (0, 1/2]$. There exists $\rho = \rho_{\eps, \delta, \beta} > 0$ such that if $\p(n) \in [\beta, 1-\beta]^{\binom{n}{2}}$, then for sufficiently large $n$, $G \sim \mathscr{G}(n + \delta, \p)$ and independent $S \sim \CU(V^{(2)})^{\otimes \eps}$, $\ev_\delta(G, S)$ holds with failure probability at most $\exp(-\rho n)$.
\end{lemma}

\noindent Here and in the remaining results, it makes no difference whether we draw $\eps = \eps(n)$ or $\eps = \eps(n+\delta)$ pairs. For simplicity in the proofs, however, we take $\eps = \eps(n)$ and require that its image be contained in $\N$. A crucial difference between the cases where $\beta = o(1)$ and $\beta = \Theta(1)$ is the adaptivity of our constants. In the former case, we can choose $\alpha$ to scale the convergence rate by any $\rho$ of our choosing. This property will be useful when we examine downstream implications.

After some simple observations on the front end, our proof of \cref{lem:bollobas-generalization} closely follows that of \cite{bollobas1990almost}. Roughly, we upper bounds the probability of any single isomorphism $\varphi$ by the probability that its orbits on $V^{(2)}$ consist entirely of edges or entirely of nonedges. We find that we can fit in additional cross-terms that account for the edits contributed by $S$. Here and in the sequel, proofs for $\beta \in \Theta(1)$ are entirely similar to those for $\beta \in o(1)$; we leave such proofs to \cref{sec:omitted-proofs}.

\begin{proof}[Proof of \cref{lem:bollobas-generalization}]
    Fix $n, \delta$. The number of choices for $W \subset V$ is \[\binom{n+\delta}{\delta} \leq 2 n^\delta,\] for $n \geq 2$. Fix one such $W$. Let $S_n^{(m)}$ denote the set of injections $W \to V$ that move exactly $m$ vertices of $W$. We have \[|S_n^{(m)}| \leq \binom{n}{m}(m+\delta)_m = n_m\binom{m + \delta}{\delta} \leq 2 n^m m^\delta,\] when $m \geq 2$, and a bound of $2\delta n$ when $m = 1$. For a map $\varphi \in S_n^{(m)}$, let $\mathscr{G} \times V^{(2)}(\varphi)$ denote the set of $(G, S)$ pairs for which $\varphi$ induces an isomorphism $G'[W] \to G'[\varphi(W)]$ for some $G' \in B_S(G)$. We let $\Pr[\varphi] = \Pr[\mathscr{G} \times V^{(2)}(\varphi)]$. Note that $\varphi$ defines a partition of $W^{(2)} \cup \varphi(W)^{(2)}$ into orbits. Let $M_i$ denote the number of orbits of order $i$, and $N = \binom{n}{2}$, so that \[\sum_{i=1}^{N} i M_i = |W^{(2)} \cup \varphi(W)^{(2)}| \geq N.\] 
    
    Note that $\varphi$ induces an isomorphism on some $G'[W] \to G'[\varphi(W)]$ if and only if each orbit consists entirely of edges or nonedges in $G'$. To bound the probability that this occurs for any $G' \in B_S(G)$, we note that for any orbit of length $i \geq 2$, we can split the orbit into $\floor{i/2}$ disjoint pairs in $V^{(2, 2)}$. For any single pair $\{xy, \vp(xy)\}$, let $p_{xy}$ and $p_{\vp(xy)}$ denote the probabilities that $xy$ and $\vp(xy)$, respectively, are included in $G$. Note that the probability $\exists G' \in B_S(G)$ such that $xy, \vp(xy) \in E(G')$ or $xy, \vp(x,y) \notin E(G')$ is bounded by \begin{align*}
        p_{xy}p_{\vp(xy)} + (1-p_{xy})(1-p_{\vp(xy)}) + \Big(p_{xy}(1-p_{\vp(xy)}) + (1-p_{xy})p_{\vp(xy)}\Big)\left(1- \left(1 - \frac{2}{\binom{n}{2}}\right)^{\eps}\right).
    \end{align*}
    Take $c > 0$ such that $\eps(n) \leq c\binom{n}{2}$ asymptotically. Hence, for any $c_0 \in (1-e^{-2c}, 1)$, sufficiently large $n$ and \cref{lem:probability-term-upper-bound} give the bound \begin{align*}
        (\alpha\beta)^2 + (1-\alpha\beta)^2 + 2\alpha\beta(1-\alpha\beta)c_0.
    \end{align*}
    Moreover, by \cref{lem:main-event-NA}, all $\sum_{i=2}^{N} M_i \floor{\frac{i}{2}}$ such events are negatively associated. Hence,
    \begin{align*}
        \Pr[\varphi] &\leq \prod_{i=2}^{N} \left( \prod_{j=1}^{\floor{\frac{i}{2}}} \Big((\alpha\beta)^2 + (1-\alpha\beta)^2 + 2\alpha\beta(1-\alpha\beta)c_0 \Big) \right)^{M_i}.
    \end{align*}
    For any $c_1 \in (0, 2-2c_0)$, since $\beta \in o(1)$, \begin{align*}
        (\alpha\beta(n))^2 + (1-\alpha\beta(n))^2 + 2\alpha\beta(n)(1-\alpha\beta(n))c_0 &= (2-2c_0)(\alpha\beta(n))^2 + (2c_0 - 2)(\alpha \beta(n)) + 1\\
        &\leq -c_1 \alpha \beta(n) + 1.
    \end{align*}
    Since $\ln(1 + x) \leq x$, we have
    \begin{align*}
        \Pr[\varphi] \leq \prod_{i=2}^{N} (-c_1 \alpha \beta(n) + 1)^{\floor{\frac{i}{2}} M_i} \leq \exp\left[ -c_1 \alpha \beta(n) \sum_{i=2}^{N} \floor{\frac{i}{2}}M_i \right].
    \end{align*}

    When $m = 1$, we have $M_1 = \binom{n-1}{2}$, $M_2 = n-1$, and $M_i = 0$ for $i \geq 3$. In this case, \[\sum_{i=2}^{N} iM_i = 2(n-1).\] When $m \geq 2$, we estimate the sum by upper bounding $M_1$. If $\varphi(xy) = xy$ but $\varphi(x) \neq x$, then $\varphi(y) = x$. Hence, for any vertex $w \in W$, $w \neq y$, we must have $\vp(xw) \neq xw$. So $M_1 \leq \binom{n-m}{2} + m/2$ and for $m \geq 2$, \[\sum_{i=2}^N iM_i \geq \binom{n}{2} - \binom{n-m}{2} - \frac{m}{2} = m\left(n - \frac{m}{2} - 1\right).\]

    Since $i \geq 2$, $\floor{\frac{i}{2}} \geq \frac{i}{4}$. We now have an upper bound of \begin{align*}
        \Pr[\varphi] &\leq \exp\left[ -\frac{c_1 \alpha}{4} \beta(n) \sum_{i=2}^{N} i M_i \right] \leq  \begin{cases}
            \exp\left[ -\frac{c_1 \alpha}{2} \beta(n)(n-1)\right] &m = 1\\
            \exp\left[ -\frac{c_1 \alpha}{4} \beta(n) m \left(n - \frac{m}{2} - 1\right)\right] &m\geq 2.
        \end{cases}
    \end{align*}
    For any $c_2 \in (0, 1)$ sufficiently large $n$ now gives \begin{align*}
        \Pr[\varphi] &\leq \begin{cases}
            \exp\left[ -\frac{c_1 c_2 
            \alpha}{2} \beta(n)n\right] &m = 1\\
            \exp\left[ -\frac{c_1 c_2 \alpha}{4} \beta(n)m n\right] &m \leq \floor{n^{1/2}}\\
            \exp\left[ -\frac{c_1 c_2 \alpha}{8} \beta(n) mn \right] &m \geq \floor{n^{1/2}}.
        \end{cases}
    \end{align*}
    Taking a union bound,
    \begin{align}
        \Pr[\ev_\delta(G, S)^c] &\leq \sum_{|W| = n} \sum_{m=1}^n \sum_{\varphi \in S_n^{(m)}} \Pr[\varphi]\nonumber\\
        &\leq 4\delta n^{\delta + 1} \exp\left[ -\frac{c_1 c_2 \alpha}{2} \beta(n)n \right] \label{eq:term-1}\\
        &\hspace{3em}+ 2n^\delta \sum_{m=2}^{\floor{n^{1/2}}} 2n^m m^\delta \exp\left[ -\frac{c_1 c_2 \alpha}{4} \beta(n)mn \right] \label{eq:term-2}\\
        &\hspace{3em}+ 2n^\delta \sum_{m=\floor{n^{1/2}}}^{n} 2n^m m^\delta \exp\left[ -\frac{c_1 c_2 \alpha}{8} \beta(n)mn \right]\label{eq:term-3}
    \end{align}
    We bound terms (\ref{eq:term-1})-(\ref{eq:term-3}) by taking $n$ sufficiently large (independent of $m$). Let $\rho' > \rho$. Choose $\alpha > \left(\frac{2}{c_1 c_2}\right)(\rho' + \delta + 1)$. \[(\ref{eq:term-1}) = \exp\left[\ln(4\delta) + (\delta + 1) \ln(n) - \frac{c_1 c_2 \alpha}{2} \beta(n)n\right] \leq \exp(-\rho' \beta(n) n).\] Next, choose $\alpha > \left(\frac{4}{c_1 c_2}\right) \left( \frac{5}{4} + \frac{3\delta}{4} + \frac{\rho'}{2}\right)$. Taking $\Delta := \frac{c_1 c_2 \alpha}{4} - \frac{3\delta}{4} - \frac{\rho'}{2} - \frac{5}{4} > 0$, we have \begin{align*}
        (\ref{eq:term-2}) &\leq \sum_{m=2}^{\floor{n^{1/2}}} \exp\left[ \ln(4) + (m + \delta)\ln(n) + \frac{\delta}{2}\ln(n) - \frac{c_1 c_2 \alpha}{4} \beta(n) m n \right]\\
        &\leq \sum_{m=2}^{\floor{n^{1/2}}} \exp\left[ m \left(\ln(2) + \left( 1 + \frac{3\delta}{4}\right)\ln(n) - \frac{c_1 c_2 \alpha}{4} \beta(n) n \right) \right]\\
        &\leq \sum_{m=2}^{\floor{n^{1/2}}} \exp\left[ m \left(\ln(2) -\Delta \beta(n)n - \left(\frac{\rho'}{2} + \frac{1}{4}\right)\beta(n)n \right) \right]\\
        &\leq \sum_{m=2}^{\floor{n^{1/2}}} \exp\left[ - \left(\rho' + \frac{1}{2}\right)\beta(n)n \right].
    \end{align*}
    Hence, \[(\ref{eq:term-2}) \leq \exp\left[ \frac{1}{2}\ln(n) - \frac{1}{2}\beta(n)n - \rho' \beta(n)n\right] \leq \exp(-\rho'\beta(n)n).\]
    Finally, choose $\alpha > \left(\frac{8}{c_1 c_2}\right)\left(\frac{3}{2} + \delta + \frac{\rho'}{2}\right)$. Taking $\Delta := \frac{c_1 c_2 \alpha}{8} - \delta - \frac{\rho'}{2} - \frac{3}{2} > 0$, we have \begin{align*}
        (\ref{eq:term-3}) &\leq \sum_{m=\floor{n^{1/2}}}^{n} \exp\left[\ln(4) + (m+2\delta)\ln(n) - \frac{c_1 c_2 \alpha}{8} \beta(n) mn \right]\\
        &\leq \sum_{m=\floor{n^{1/2}}}^{n} \exp\left[m\left(\ln(2) + \left(1 + \delta\right)\ln(n) - \frac{c_1 c_2 \alpha}{8} \beta(n) n \right)\right]\\
        &\leq \sum_{m=\floor{n^{1/2}}}^{n} \exp\left[m\left(\ln(2) - \Delta \beta(n)n - \left(\frac{\rho'}{2} + \frac{1}{2}\right)\beta(n)n \right)\right].
    \end{align*}
    Hence, \[(\ref{eq:term-3}) \leq \exp\left[\ln(n) - (\rho' + 1)\beta(n)n\right] \leq \exp(-\rho' \beta(n)n).\]
    Combining terms, \[\Pr[\ev_\delta(G,S)^c] \leq 3\exp(-\rho' \beta(n)n) \leq \exp(-\rho \beta(n)n).\]
\end{proof}

Drawing $S \sim \CU(V^{(2)})^{\otimes \eps}$ is convenient for analysis but perhaps unnatural. This choice is without any loss of expressive power, however. We can use \cref{lem:bollobas-generalization,lem:bollobas-generalization-constant-prob} in a black-box fashion to obtain results for $S \sim \CU(V^{(2, \eps)})$. 

\begin{theorem}\label{thm:bollobas-generalization-wo-replacement}
    Let $\delta \in \N$, $\rho > 0$. Let $\eps : \N \to \N$ be such that $\eps(n) \leq c\binom{n}{2}$ for some $c \in (0,1)$. Then there exists some $\alpha = \alpha_{c, \delta, \rho} > 0$ such that if \[\frac{\ln(n)}{n} \leq \beta(n), \ \ \beta(n) \in o(1),\] and $\p(n) \in [\alpha \beta(n), 1-\alpha \beta(n)]^{\binom{n}{2}}$, then for sufficiently large $n$, $G \sim \mathscr{G}(n + \delta, \p)$ and independent $S \sim \CU(V^{(2, \eps)})$, $\ev_\delta(G, S)$ holds with failure probability  $\exp(-\rho \beta(n)n)$.
\end{theorem}

\begin{theorem}\label{thm:bollobas-generalization-wo-replacement-constant-prob}
    Let $\delta \in \N$, $\beta \in (0, 1/2]$. Let $\eps : \N \to \N$ be such that $\eps(n) \leq c\binom{n}{2}$ for some $c \in (0,1)$. There exists some $\rho = \rho_{c, \delta, \beta} > 0$ such that if $\p(n) \in [\beta,1-\beta]^{\binom{n}{2}}$, then for sufficiently large $n$, $G \sim \SG(n+\delta, \p)$ and independent $S \sim \CU(V^{(2, \eps)})$, $\ev_\delta(G, S)$ holds with failure probability at most $\exp(-\rho n)$.
\end{theorem}

\begin{proof}[Proof of \cref{thm:bollobas-generalization-wo-replacement}]
Let $\rho' > \rho$, $\eps(n) \leq c \binom n2$, and abbreviate $N = \binom n2$ for concision. Then set
\[ \eps'(n) = N \log \left( \frac{1}{1 - \gamma \eps(n) / N} \right), \] 
with $\gamma = \frac{1 + 1/c}{2}$. It is not difficult to see that $\eps'(n) = O(N) = O(n^2)$, as $\gamma \eps(n) / N \leq \gamma \cdot c = \frac{c + 1}{2} < 1$. Now consider the random variable $Z_n = |S|$ for $S \sim \CU(V^{(2)})^{\otimes \eps'(n)}$. Equivalently, $Z_n = \sum_{1}^N X_i$ where $X_i = \ind{i \in S}$. We start by showing $Z_n \geq \eps(n)$ with probability $\Omega(1)$. 

To this end, first note that $\E(Z_n)$ exceeds $\eps(n)$ by a constant factor: 
\begin{align*}
\E(Z_n) &= \sum_{i \in [N]} \E(X_i) \\ 
&= N \cdot \left(1 - \left( 1 - \frac 1N \right)^{\eps'(n)} \right) \\ 
&\geq N \left(1 - \exp(- \eps'(n) / N) \right) \\
&= N \Big( 1 - \exp \big(- \log((1 - \gamma \eps'(n) / N)^{-1}) \big) \Big) \\ 
&= N\big(1 - (1 - \gamma \eps(n) / N)\big) \\
&= \gamma \eps(n). 
\end{align*}
The inequality simply leverages that $1 + x \leq \exp(x)$. We now bound $\Var(Z_n)$ from above by $\E(Z_n)$, by appealing to \cref{lem:S-containment-NA}. Letting $p := 1 - (1 - 1/N)^{\eps'(n)}$, the sub-additivity of variance for negatively associated variables gives
\begin{align*}
\Var(Z_n) &\leq \sum_{i=1}^N \Var(X_i) = N \cdot p(1 - p) \leq N p = \E(Z_n). 
\end{align*}
Finally, we invoke the Paley-Zygmund inequality with $\theta = 1 / \gamma < 1$. This yields
\begin{align*} 
\P\Big(Z_n \geq \eps(n) \Big) &\geq \P\Big( Z_n \geq \frac{1}{\gamma} \cdot \E(Z_n) \Big) \\
&\geq \left(1 - \frac 1 \gamma \right)^2 \frac{\E(Z_n)^2}{\E(Z_n^2)} \\ 
&= \left(1 - \frac 1 \gamma \right)^2 \frac{\E(Z_n)^2}{\E(Z_n)^2 + \Var(Z_n)} \\
&\geq \frac{(1 - 1/\gamma)^2}{1 + 1 / \E(Z_n)} \\
&\geq \frac{(1 - 1/\gamma)^2}{2} \\
&= \Omega(1),
\end{align*}
as desired. The penultimate line uses that $\E(Z_n) \geq 1$, and the previous line uses $\Var(Z_n) \leq \E(Z_n)$. 
By \cref{lem:bollobas-generalization} with $\eps'(n)$, we now have \begin{align}\label{eq:eps-prime-bound}
    \Pr_{\substack{G \sim \SG(n+\delta, \p)\\S \sim (V^{(2)})^{\otimes \eps'}}}\left[E_\delta(G, S)^c \, \big| \, |S| \geq \eps\right] \leq O(\exp(-\rho' \beta(n) n)),
\end{align}
and conclude by conditioning. To be explicit, note that for fixed $G$ and $1 \leq k \leq \eps' - 1$, \begin{align*}
    \Pr_{S \sim (V^{(2)})^{\otimes \eps'}} \left[E_\delta(G, S)^c \, \big| \, |S| = k \right] &= \Pr_{S \sim V^{(2, k)}}[ E_\delta(G, S)^c]\\
    &= \sum_{|S| = k} \Pr_{S \sim V^{(2, k)}}[S] \cdot \ind{E_\delta(G, S)^c}\\
    &\leq \sum_{|S| = k} \Pr_{S \sim V^{(2, k)}}[S] \cdot \Pr_{e \sim V\setminus S}[E_\delta(G, S \cup \{e\})^c]\\
    &= \Pr_{\substack{S \sim V^{(2, k)}\\e \sim V^{(2)}\setminus S}}[E_\delta(G, S \cup \{e\})^c]\\
    &= \Pr_{\substack{S \sim (V^{(2)})^{\otimes \eps'}}} \left[E_\delta(G, S)^c \, \big| \, |S| = k + 1 \right].
\end{align*}
Expanding the left hand side of \cref{eq:eps-prime-bound} and invoking the independence of $G$ and $S$, we obtain failure probability $\eps(-\rho \beta(n)n)$ for $S \sim V^{(2, \eps)}$.
\end{proof}

\noindent \cref{thm:bollobas-generalization-wo-replacement-constant-prob} follows from  a similar argument, instead making a call to \cref{lem:bollobas-generalization-constant-prob}.

\subsection{Planted Structure}\label{sec:planted-structure}

Modulating $\eps(n)$ appropriately, the results in \cref{sec:robust-asymmetry} immediately imply the preservation of subgraph asymmetry properties under the semi-random interpretation. We address two regimes, determined by whether an adversary plants structure before or after $G$ is sampled.

\begin{corollary}\label{cor:wo-replacement-small-p-planted}
    Let $\delta \in \N$, $B > 0$, $\rho > 0$. There exists some $\alpha = \alpha_{\delta, B, \rho} > 0$ such that if \[\frac{\ln(n)}{n} \leq \beta(n), \ \ \beta(n) \in o(1),\] and $\p(n) \in [\alpha\beta(n), 1-\alpha\beta(n)]^{\binom{n}{2}}$, then for sufficiently large $n$, $G \sim \SG(n+\delta, \p)$ and any fixed $P \subset V^{(2)}$ such that $|P| \leq B\beta(n)n$, $\ev_\delta(B_P(G))$ fails with probability at most $\exp(-\rho\beta(n)n)$. In particular, for any partition $P_1, P_2$ of $P$, $\ev_\delta(G_{P_1, P_2})$ fails with probability at most $\exp(-\rho\beta(n)n)$.
\end{corollary}

\begin{corollary}\label{cor:wo-replacement-constant-p-planted}
    Let $\delta \in \N$, $\beta \in (0, 1/2]$. There exists some $\rho = \rho_{\delta, \beta} > 0$ and $B = B_{\delta, \beta} > 0$ such that if $\p(n) \in [\beta, 1-\beta]^{\binom{n}{2}}$, then for sufficiently large $n$, $G \sim \SG(n+\delta, \p)$ and any fixed $P \subset V^{(2)}$ such that $|P| \leq Bn$, $\ev_\delta(B_P(G))$ fails with probability at most $\exp(-\rho n)$. In particular, for any partition $P_1, P_2$ of $P$, $\ev_\delta(G + P_1 - P_2)$ fails with probability at most $\exp(-\rho n)$.
\end{corollary}

\begin{proof}[Proof of \cref{cor:wo-replacement-small-p-planted}]
    Take $0 < c_0 < c_1 < c_2 < 1$. Let $S \sim \CU(V^{(2, \eps)})$ independently, for $c_1\binom{n}{2} \leq \eps(n) \leq \floor{c_2\binom{n}{2}}$. Also take $\rho' = \rho + B\ln(1/c_0)$. By \cref{thm:bollobas-generalization-wo-replacement}, we can take $\alpha = \alpha_{c_2, \delta, \rho'} > 0$ such that for large enough $n$, $\ev_\delta(G, S)$ fails with probability at most $\exp(-\rho'\beta(n)n)$. 

    Note that $|P| \leq \binom{n}{2}$. Hence, for any $S' \subset V^{(2)}$ satisfying $P \subset S'$, \[\Pr_G[\ev_\delta(B_P(G))^c] \leq \Pr_G[\ev_\delta(G, S')^c].\] Since $|P| \leq \eps$, $\Pr_S[P\subset S] > 0$ and the law of total probability yields \[\Pr_G[\ev_\delta(B_P(G))^c] \leq \left(\Pr_S[P \subset S]\right)^{-1} \Pr_{G, S}[\ev_\delta(G, S)^c].\]
    By assumption, $\beta(n)n \in o(n) \subset o(n^2)$, so \cref{lem:random-subset-exponential-lower-bound} gives \[\Pr_S[P \subset S] \geq \left(\frac{c_1 \binom{n}{2} - B\beta(n)n}{\binom{n}{2}}\right)^{|P|} \geq \exp[-\ln(1/c_0)B\beta(n)n].\] Hence, \[\Pr_G[\ev_\delta(B_P(G))^c] \leq \exp\Big[(B\ln(1/c_0) - \rho')\beta(n)n\Big] = \exp(-\rho \beta(n)n).\]
\end{proof}

Note that in the proof of \cref{cor:wo-replacement-small-p-planted}, we make use the adaptive constants in \cref{thm:bollobas-generalization-wo-replacement}. We can again use this adaptivity to recover variations on existing results in the literature on property resilience \citep{sudakov2008local,frieze2015introduction,bollobas2011random}.

\begin{corollary}\label{cor:wo-replacement-small-p-ball}
    Let $\delta \in \N$, $\rho > 0$, and \[\frac{\ln(n)}{n} \leq \beta(n), \ \ \beta(n) \in o(1).\] Also let $\eps \in O(\beta(n)n/\log(n))$. There exists some $\alpha = \alpha_{\eps, \delta, \rho} > 0$ such that if $\p(n) \in [\alpha\beta(n), 1-\alpha\beta(n)]^{\binom{n}{2}}$, then for sufficiently large $n$ and $G \sim \SG(n+\delta, \p)$, $\ev_\delta(B_\eps(G))$ fails with probability at most $\exp(-\rho \beta(n)n)$.
\end{corollary}

\begin{corollary}\label{cor:wo-replacement-constant-p-ball}
    Let $\delta \in \N$, $\beta \in (0, 1/2]$. There exists some $\rho = \rho_{\delta, \beta} > 0$ and $B  = B_{\delta,\beta} > 0$ such that if $\p(n) \in [\beta, 1-\beta]^{\binom{n}{2}}$ and $\eps(n) \leq Bn/\log(n)$, then for sufficiently large $n$ and $G \sim \SG(n+\delta, \p)$, $\ev_\delta(B_\eps(G))$ fails with probability at most $\exp(-\rho n)$.
\end{corollary}

\begin{proof}[Proof of \cref{cor:wo-replacement-small-p-ball}]
    There exists $B > 0$ such that for sufficiently large $n$, $\eps(n) \leq B\beta(n)n/\log(n)$. Let $c_0 > 0$ and $\rho' = \rho + (2+c_0)B$. By \cref{cor:wo-replacement-small-p-planted}, there exists $\alpha = \alpha_{\eps, \delta, \rho'}$ so that for sufficiently large $n$ and any fixed $P$ satisfying $|P| \leq B\beta(n)n$, $E_\delta(B_P(G))$ fails with probability at most $\exp(-\rho'n)$. 
    Hence, \begin{align*}
        \Pr_G[\ev_\delta(B_\eps(G))^c] = \Pr_G \left[ \bigvee_{|P| = \eps} \ev_\delta(B_P(G))^c \right] \leq \binom{\binom{n}{2}}{\eps} \exp(-\rho' \beta(n)n).
    \end{align*}
    By the monotonicity of $\left(\frac{em}{x}\right)^x$ on $(0, m]$, \begin{align*}
        \log\binom{\binom{n}{2}}{\eps} \leq \left(\frac{B\beta(n)n}{\log(n)}\right)\left(\log\left(\frac{e}{2B}\right) + \log(n) + \log\left(\frac{\log(n)}{\beta(n)}\right)\right) \leq (2+c_0) B\beta(n)n.
    \end{align*}
\end{proof}

\noindent For constant $|\p|_\infty$, these results confirm the preservation of symmetry even if an adversary plants a clique on $\Theta(\sqrt{n/\log(n)})$ vertices after seeing the sampled graph.

\subsection{Reconstruction Consequences}

Through \cref{lem:n-2-anch-generalization}, a number of reconstruction results follow immediately from \cref{thm:bollobas-generalization-wo-replacement-informal,cor:wo-replacement-planted-informal,cor:wo-replacement-p-ball-informal}. Using techniques from \cite{bollobas1990almost}, one can even show that semi-random graphs have reconstruction number three with failure probability $\exp(-\Omega(\beta(n)n))$. Perhaps more interesting is the following characterization of nonreconstructible graphs, which follows from \cref{cor:wo-replacement-small-p-ball,cor:wo-replacement-constant-p-ball}.

\begin{corollary}\label{cor:nonrecon-balls-small-p}
    Let $\beta(n) \in o(1)$ such that $\log(n)/n \leq \beta(n)$, and let $\eps \in O(\beta(n)n/\log(n))$. Then there exists $\alpha > 0$ such that if $\p(n) \in [\alpha\beta(n),1-\alpha\beta(n)]^{\binom{n}{2}}$, then \[\Pr_{\SG(n, \p)}[B_\eps(\ol{\mathfrak{R}}_n)] \xrightarrow{n\to\infty} 0.\]
\end{corollary}

\begin{corollary}\label{cor:nonrecon-balls-constant-p}
    Let $\beta \in (0, 1/2]$. There exists $B > 0$ such that if $\eps(n) \leq Bn/\log(n)$ and $\p(n) \in [\beta, 1-\beta]^{\binom{n}{2}}$, then \[\Pr_{\SG(n, \p)}[B_\eps(\ol{\mathfrak{R}}_n)] \xrightarrow{n\to\infty} 0.\]
\end{corollary}

\section{Outlook}

As noted in the overview, when $S \sim \CU(V^{(2)})^{\otimes \eps}$, any reconstructibility result holding with probability $1-o(1)$ and $\eps = \Omega(n^2\log(n))$ is equivalent to the deterministic conjecture for large $n$. This leaves a logarithmic gap unknown. Similarly, when $S \sim \CU(V^{(2,\eps)})$, it remains unclear whether one can handle any $\eps = n^2/2 - o(n)$. Such questions outline what is possibly the most natural line for further inquiry.

As suggested in \cref{sec:primary-results}, one can view the semi-random extension of the smoothed model as, instead of uniformly widening the hypercube padding on $\p \in [0,1]^N$, allowing a random (or adversarially chosen) subset $S$ of the entries in this tuple to be arbitrary. One might consider a relaxed setting, where instead of letting $\{p_i\}_{i\in S}$ be arbitrary, widening the bound on these entries beyond the $\Theta(\log(n)/n)$ barrier. One may hope to successfully do this on $n^2/2 - o(n)$ of the entries.

Beyond the semi-random model, there exist several natural notions of approximate graph reconstruction. For instance, take $\operatorname{d\mathfrak{R}}(G)$ to be the minimum number of edge additions/removals necessary to make $G$ reconstructible. One can always isolate a vertex to make $G$ reconstructible, giving $\operatorname{d\mathfrak{R}}(G) \leq n-1$. To the best of our knowledge, however, no sublinear bounds have been established. Interestingly, by strategically changing the degree profile and using \citet[Theorem~12]{myrvold1987bidegreed}, one can show that only three edits suffice to make any graph on at least four vertices edge-reconstructible—reconstructible from its multiset of subgraphs with a single edge removed \citep{bondy1991manual}. We leave examination of these bounds to future work.

\newpage 
\bibliographystyle{plainnat-nolinks.bst}
\bibliography{refs.bib}
\newpage 

\appendix
\section{Probability}
\label{sec:probability-tools}

We first record a simple inequality that enables the most crucial step in our analysis.

\begin{lemma}\label{lem:probability-term-upper-bound}
    Let $a \in (0,1/2]$, $c_0 \in (0, 1)$. Define \begin{align*}
        f : [a,1-a]^2 &\to [0,1]\\
        (x, y) &\mapsto xy + (1-x)(1-y) + (x(1-y) + (1-x)y)c_0.
    \end{align*}
    Then for all $x, y \in [a, 1-a]$, $f(x, y) = f(1-x, 1-y)$ and $f(x, y) \leq f(a, a)$.
\end{lemma}

\begin{proof}[Proof of \cref{lem:probability-term-upper-bound}]
    Since $f(x,y) = f(1-x, 1-y)$, we may assume that $x \geq 1/2$. \begin{align*}
        f(x, y) &= (2-2c_0)xy + (c_0 - 1)x + (c_0 - 1)y + 1\\
        &= [(2-2c_0)x - (1-c_0)]y + (c_0 - 1)x + 1\\
        &\leq [(2-2c_0)x-(1-c_0)](1-a) + (c_0 - 1)x + 1\\
        &= [(2-2c_0)(1-a)-(1-c_0)]x - (1-c_0)(1-a) + 1\\
        &\leq [(2-2c_0)(1-a)-(1-c_0)](1-a) - (1-c_0)(1-a) + 1\\
        &= (2-2c_0)(1-a)^2 - (2-2c_0)(1-a) + 1\\
        &= f(1-a, 1-a) = f(a, a). \qedhere
    \end{align*}
\end{proof}

\noindent The following inequality will also be useful.

\begin{lemma}\label{lem:random-subset-exponential-lower-bound}
    Let $T \subset U$ such that $|T| = k$, $|U| = n$, and $n \geq m \geq k$. Then for $S \sim \CU(U^{(m)})$, \[\Pr_S[T \subset S] \geq \left(\frac{m-k}{n}\right)^k.\]
\end{lemma}

\begin{proof}
    To see this, note that if $T = \{x_1, \ldots, x_k\}$, then for $i \in [k]$, \[\Pr[x_i \in S | x_1, \ldots, x_{i-1} \in S] = \frac{\binom{n-(i-1)-1}{m-(i-1)-1}}{\binom{n-(i-1)}{m-(i-1)}} = \frac{m-i+1}{n-i+1}.\] By chain rule, \begin{align*}
        \Pr_S[T \subset S] = \prod_{j=1}^{k} \left(\frac{m-i+1}{n-i+1}\right) \geq \left(\frac{m-k}{n}\right)^k. 
    \end{align*} 
\end{proof}

\subsection{Negative Association}

Crucial in our analysis are some basic facts about negative association. For brevity, we say that a set of random variables is NA if it is negatively associated.

\begin{lemma}\label{lem:S-containment-NA}
    For $S \sim \CU(V^{(2)})^{\otimes \eps}$, let $\ind{e \in S}$ denote the random variable that takes on the value 1 when $e = S_j$ for some $j$ and 0 otherwise. Then $\{\ind{e \in S}\}_{e \in V^{(2)}}$ is NA. In particular, for any $F \subset V^{(2)}$, \[\Pr[\bigwedge_{e \in F} e \in S] \leq \prod_{e \in F} \Pr[e \in S].\]
\end{lemma}

\begin{proof}
    We closely follow a standard proof for balls and bins. 
    For $e \in V^{(2)}$ and $j \in [\eps]$, let $X_{e, j}$ be a random variable taking the value 1 if $S_j = e$ and 0 otherwise. By the Zero-One Lemma \citep{dubhashi1998balls},
    $\{X_{e, j}\}_e$ is NA. Since unions of mutually independent NA families are NA, $\{X_{e, j}\}_{e, j}$ is NA \citep{joag-dev1983negative}. But note that $\{\ind{e \in S}\}_{e \in V^{(2)}}$ are all monotone increasing in $\{X_{e, j}\}_{e, j}$ and defined on disjoint subsets of $\{X_{e, j}\}_{e, j}$, hence NA. Using the marginal probability bound for NA families, \[\Pr[\bigwedge_{e \in F} e \in S] = \Pr[\bigwedge_{e \in F} \ind{e \in S} \geq 1] \leq \prod_{e \in F} \Pr[\ind{e \in S} \geq 1] = \prod_{e \in F} \Pr[e \in S].\]
\end{proof}

\begin{lemma}\label{lem:main-event-NA}
    Let $p \in (0, 1)$, $\eps \in \N$, $G \sim \mathscr{G}(n, p)$ and independent $S \sim \mathcal{U}(V^{(2)})^{\otimes \eps}$. Let $A, B \subset V^{(2)}$ be disjoint, $A = \{e_j\}_1^k$, $B = \{f_j\}_1^k$. For $j \in [k]$, let $E_j$ be the event given by \[(e_j \in E \land f_j \in E ) \lor (e_j \notin E \land f_j \notin E) \lor \Big[ \Big( (e_j \in E \land f_j \notin E) \lor (e_j \notin E \land f_j \in E) \Big) \land (e_j \in S \lor f_j \in S) \Big].\] Then $\{\ind{E_j}\}_j$ is NA. In particular, \[\Pr[\bigwedge_j E_j] \leq \prod_{j} \Pr[E_j].\]
\end{lemma}

\begin{proof}
    For $j \in [k]$, define $X_j$ as the indicator of the event \[(e_j \in E \land f_j \in E ) \lor (e_j \notin E \land f_j \notin E).\] For $e \in V^{(2)}$, also define $Y_e = \ind{e \in S}$. Note that $\{X_j\}_j$ are mutually independent, and that $\{Y_e\}_{e \in A \cup B}$ is NA by \cref{lem:S-containment-NA}. Since these collections are independent, $\CA = \{X_j\}_j \cup \{Y_e\}_{e \in A \cup B}$ is NA \citep{joag-dev1983negative}. Now observe that \[\ind{E_j} = X_j + (1-X_j)\ind{Y_{e_j} + Y_{f_j} \geq 1}.\] So $\{\ind{E_j}\}_j$ are monotonically increasing in $\CA$ and defined on disjoint subsets of $\CA$, and hence NA. Using the marginal probability bound for NA families, \[\Pr[\bigwedge_j E_j] = \Pr[\bigwedge_j \ind{E_j} \geq 1] \leq \prod_j \Pr[\ind{E_j} \geq 1] = \prod_{j} \Pr[E_j].\]
\end{proof}

\section{Omitted Proofs}
\label{sec:omitted-proofs}

\subsection{Proof of \cref{lem:bollobas-generalization-constant-prob}}

\begin{proof}[Proof of \cref{lem:bollobas-generalization-constant-prob}]
    We follow the proof of \cref{lem:bollobas-generalization}, except for minor changes. Take $c > 0$ such $\eps(n) \leq c\binom{n}{2}$ asymptotically. For any $c_0 \in (1-e^{-2c}, 1)$, \cref{lem:probability-term-upper-bound} and \cref{lem:main-event-NA} give
    \begin{align*}
        \Pr[\varphi] &\leq \prod_{i=2}^{N} \left( \prod_{j=1}^{\floor{\frac{i}{2}}} \Big(\beta^2 + (1-\beta)^2 + 2\beta(1-\beta)c_0 \Big) \right)^{M_i}.
    \end{align*}
    Note that $\beta^2 + (1-\beta)^2 + 2\beta(1-\beta)c_0 < 1$. Taking $c_1 = -\ln(\beta^2 + (1-\beta)^2 + 2\beta(1-\beta)c_0) > 0$,
    \begin{align*}
        \Pr[\varphi] \leq \exp\left[ - c_1 \sum_{i=2}^{N} \floor{\frac{i}{2}}M_i \right].
    \end{align*}
    As before, we have \begin{align*}
        \Pr[\varphi] &\leq \exp\left[ -\frac{c_1}{4} \sum_{i=2}^{N} i M_i \right] \leq  \begin{cases}
            \exp\left[ -\frac{c_1}{2} (n-1)\right] &m = 1\\
            \exp\left[ -\frac{c_1}{4} m \left(n - \frac{m}{2} - 1\right)\right] &m\geq 2.
        \end{cases}
    \end{align*}
    For any $c_2 \in (0, 1)$, sufficiently large $n$ gives \begin{align*}
        \Pr[\varphi] &\leq \begin{cases}
            \exp\left[ -\frac{c_1 c_2}{2} n\right] &m = 1\\
            \exp\left[ -\frac{c_1 c_2}{4} m n\right] &m \leq \floor{n^{1/2}}\\
            \exp\left[ -\frac{c_1 c_2}{8} mn \right] &m \geq \floor{n^{1/2}}.
        \end{cases}
    \end{align*}
    Taking a union bound, \begin{align}
        \Pr[\ev_\delta(G, S)^c] &\leq \sum_{|W| = n} \sum_{m=1}^n \sum_{\varphi \in S_n^{(m)}} \Pr[\varphi]\nonumber\\
        &\leq 4\delta n^{\delta + 1} \exp\left[ -\frac{c_1 c_2}{2} n\right] \label{eq:term-1-b}\\
        &\hspace{2em}+ 2n^\delta \sum_{m=2}^{\floor{n^{1/2}}} 2n^m m^\delta \exp\left[ -\frac{c_1 c_2}{4} m n\right] \label{eq:term-2-b}\\
        &\hspace{2em}+ 2n^\delta \sum_{m=\floor{n^{1/2}}}^{n} 2n^m m^\delta \exp\left[ -\frac{c_1 c_2}{8} mn \right]. \label{eq:term-3-b}
    \end{align}
    We bound terms (\ref{eq:term-1-b})-(\ref{eq:term-3-b}). Take $0 < \rho < \rho' < \rho'' < \frac{c_1 c_2}{4}$. Since $\rho' < \frac{c_1 c_2}{2}$, it's clear that term (\ref{eq:term-1-b}) is bounded by $\exp(-\rho' n)$. Next, \begin{align*}
        (\ref{eq:term-2-b}) &\leq \sum_{m=2}^{\floor{n^{1/2}}} \exp\left[ \ln(4) + \left(\frac{3\delta}{2} + m\right)\ln(n) - \frac{c_1 c_2}{4}mn \right]\\
        &\leq \sum_{m=2}^{\floor{n^{1/2}}} \exp\left[ m \left(\ln(2) + \left(\frac{3\delta}{4} + 1\right)\ln(n) - \frac{c_1 c_2}{4}n \right) \right]\\
        &\leq \sum_{m=2}^{\floor{n^{1/2}}} \exp( -m \rho'' n) \leq \exp\left[\frac{1}{2}\ln(n) - 2\rho''n \right] \leq \exp(-2\rho' n).
    \end{align*}
    Lastly, \begin{align*}
        (\ref{eq:term-3-b}) &\leq \sum_{m=\floor{n^{1/2}}}^n \exp\left[\ln(4) + (m + 2\delta)\ln(n) - \frac{c_1 c_2}{8} mn\right]\\
        &\leq \sum_{m=\floor{n^{1/2}}}^n \exp\left[ m \left(\ln(2) + (1 + \delta)\ln(n) - \frac{c_1 c_2}{8} n \right)\right]\\
        &\leq \sum_{m=\floor{n^{1/2}}}^n \exp\left[m\left(-\frac{\rho''}{2}n\right)\right] \leq \exp(\ln(n) - \rho'' n) \leq \exp(-\rho' n).
    \end{align*}
    Combining terms, \[\Pr[\ev_\delta(G, S)^c] \leq 3 \exp(-\rho' n) \leq \exp(-\rho n).\]
\end{proof}

\subsection{Proof of \cref{cor:wo-replacement-constant-p-planted}}

\begin{proof}[Proof of \cref{cor:wo-replacement-constant-p-planted}]
    Take $0 < c_0 < c_1 < c_2 < 1$. Let $S \sim \CU(V^{(2,\eps)})$ independently, for $c_1 \binom{n}{2} \leq k \leq \floor{c_2\binom{n}{2}}$. By \cref{thm:bollobas-generalization-wo-replacement-constant-prob}, there exists some $\rho' = \rho_{c_2, \delta, \beta}' > 0$ such that for sufficiently large $n$, $\ev_\delta(G, S)$ fails with probability at most $\exp(-\rho' n)$.

    Take $B < \rho'/\log(1/c_0)$. Note that $|P| \leq \binom{n}{2}$, so for any $S' \subset V^{(2)}$ satisfying $P \subset S'$, 
    \begin{align*}
        \Pr_{G} [\ev_\delta(B_P(G))^c] &\leq \Pr_{G} [\ev_\delta(G, S')^c].
    \end{align*}
    Since $|P| \leq \eps$, $\Pr_S[P \subset S] > 0$ and the law of total probability yields \[\Pr_G[\ev_\delta(B_P(G))^c] \leq \left(\Pr_S[P \subset S]\right)^{-1} \Pr_{G, S}[\ev_\delta(G, S)^c].\]
    Moreover, \cref{lem:random-subset-exponential-lower-bound} gives \begin{align*}
        \Pr_S[P \subset S] \geq \left(\frac{c_1\binom{n}{2}-Bn}{\binom{n}{2}}\right)^{|P|} \geq \exp\left[-\ln(1/c_0) Bn\right].
    \end{align*}
    Taking $\rho = \rho' - \ln(1/c_0)B > 0$, we have
    \begin{align*}
        \Pr_{G} [\ev_\delta(B_P(G))^c] &\leq \exp(\ln(1/c_0)Bn - \rho' n) = \exp(-\rho n).
    \end{align*}
\end{proof}

\subsection{Proof of \cref{cor:wo-replacement-constant-p-ball}}

\begin{proof}[Proof of \cref{cor:wo-replacement-constant-p-ball}]
    By \cref{cor:wo-replacement-constant-p-planted}, there exists $\rho' = \rho_{\delta, \beta}' > 0$ and $B' = B_{\delta, \beta}' > 0$ so that for sufficiently large $n$ and any fixed $P$ satisfying $|P| \leq B'n$, $E_\delta(B_P(G))$ fails with probability at most $\exp(-\rho'n)$. 
    Hence, \begin{align*}
        \Pr_G[\ev_\delta(B_\eps(G))^c] = \Pr_G \left[ \bigvee_{|P| = \eps} \ev_\delta(B_P(G))^c \right] \leq \binom{\binom{n}{2}}{\eps} \exp(-\rho' n).
    \end{align*}
    Let $c_0 > 0$. By the monotonicity of $\left(\frac{em}{x}\right)^x$ on $(0, m]$, \begin{align*}
        \log\binom{\binom{n}{2}}{\eps} \leq \left(\frac{Bn}{\log(n)}\right)\left(\log\left(\frac{e}{2B}\right) + \log(n) + \log\log(n)\right) \leq (2+c_0) Bn.
    \end{align*}
    Taking $B < \min\{B', \rho'/(2+c_0) \}$ and $\rho = \rho' - (2+c_0)B$ gives the result.
\end{proof}

\section{Connection to Deterministic Reconstruction}
\label{sec:deterministic-reconstruction}

For completeness, we record connections between deterministic and probabilistic reconstruction statements. We do not pin down precise thresholds.

\begin{observation}
    Suppose there exists a family of distributions $\{\mathscr{D}_n\}_1^\infty$ over undirected graphs on $n$ vertices such that for $G \sim \mathscr{D}_n$, $S \sim (V^{(2)})^{\otimes \eps}$, and $\eps(n) \geq C n^2 \log n$ for $C > 1$, \begin{align*}
        \Pr [B_S(G) \subset \mathfrak{R}_n] = 1-o(1).
    \end{align*}
    Then the reconstruction conjecture is true for sufficiently large $n$.
\end{observation}

\begin{proof}
    Let $H$ be a graph on $n$ vertices. Hence, for $G \sim \mathscr{D}_n$, if $H \notin \mathfrak{R}$, then \[\Pr[ B_S(G) \not\subset \mathfrak{R} \, | \, S = V^{(2)}] = 1.\] Expanding, we have\begin{align*}
        \Pr[B_S(G) \not\subset \mathfrak{R}_n]  \geq \Pr[S = V^{(2)}] \geq 1-N^{1-C} \geq 1 - o(1),
    \end{align*}
    yielding a contradiction for sufficiently large $n$.
\end{proof}

\begin{observation}
    Let $\beta \in O(1/n^2)$ and $\gamma \in o(1)$. Suppose $G \sim \SG(n, \p)$ is reconstructible with probability at least $1-\gamma(n)$ for all \begin{align*}
        \p_i \in [\beta(n), 1-\beta(n)] \quad i \in \binom{n}{2}.
    \end{align*}
    Then the reconstruction conjecture is true for sufficiently large $n$.
\end{observation}

\begin{proof}
    Let $H$ be a graph $n$ vertices, and define \[\p_i := \beta(n) \cdot \ind{i \notin E(H)} + (1-\beta(n)) \cdot \ind{i \in E(H)}.\] If $H \notin \mathfrak{R}$, then \begin{align*}
        \Pr[G \notin \mathfrak{R}] \geq \Pr[G = H] = \Omega(1).
    \end{align*}
    Since the left-hand side is $o(1)$, we have a contradiction for sufficiently large $n$.
\end{proof}

\end{document}